\pdfoutput=1
\documentclass[conference]{IEEEtran}

\usepackage[pdftex]{graphicx}
\usepackage{hyperref, url} 
\usepackage{threeparttable} 
\usepackage{booktabs} 
\usepackage{amssymb} 
\usepackage{amsmath}
\usepackage{multirow} 
\usepackage{pifont} 
\newcommand{\cmark}{\ding{51}}%
\newcommand{\xmark}{\ding{55}}%
\usepackage{caption}
\usepackage{tabularx}
\usepackage[numbers,sort&compress,square]{natbib} 

\usepackage{subcaption}
\newcommand{\smalltilde}{\raise.17ex\hbox{$\scriptstyle\mathtt{\sim}$}}

\begin{document}
\title{Time-Based CAN Intrusion Detection Benchmark\thanks{\scriptsize{
This manuscript has been co-authored by UT-Battelle, LLC, under contract DE-AC05-00OR22725 with the US Department of Energy (DOE). The US government retains and the publisher, by accepting the article for publication, acknowledges that the US government retains a nonexclusive, paid-up, irrevocable, worldwide license to publish or reproduce the published form of this manuscript, or allow others to do so, for US government purposes. DOE will provide public access to these results of federally sponsored research in accordance with the DOE Public Access Plan (http://energy.gov/downloads/doe-public-access-plan).
}}
}


\author{
\IEEEauthorblockN{Deborah H. Blevins\printfnsymbol{1}\thanks{\printfnsymbol{1}Deborah H. Blevins and Pablo Moriano, placed in alphabetical order, contributed equally to this work.}\authorrefmark{2}, Pablo Moriano\printfnsymbol{1}\authorrefmark{3}, Robert A. Bridges\authorrefmark{3}, Miki E. Verma\authorrefmark{3}\\ 
Michael D. Iannacone\authorrefmark{3}, Samuel C Hollifield\authorrefmark{3}
}
\IEEEauthorblockA{\authorrefmark{2}University of Kentucky, \authorrefmark{3}Oak Ridge National Laboratory\\
deborah.blevins@uky.edu, $\{$moriano, bridgesra, vermake, iannaconemd, hollifieldsc$\}$@ornl.gov}
}


\IEEEoverridecommandlockouts
\makeatletter\def\@IEEEpubidpullup{6.5\baselineskip}\makeatother
\IEEEpubid{\parbox{\columnwidth}{
    Workshop on Automotive and Autonomous Vehicle Security (AutoSec) 2021 \\
    21 February 2021 \\
    ISBN 1-891562-68-1 \\
    https://dx.doi.org/10.14722/autosec.2021.23xxx \\
    www.ndss-symposium.org
}
\hspace{\columnsep}\makebox[\columnwidth]{}}

\makeatletter
\newcommand{\printfnsymbol}[1]{%
  \textsuperscript{\@fnsymbol{#1}}%
}

\maketitle

\begin{abstract}
Modern vehicles are complex cyber-physical systems made of hundreds of electronic control units (ECUs) that communicate over controller area networks (CANs). This inherited complexity has expanded the CAN attack surface which is vulnerable to message injection attacks. These injections change the overall timing characteristics of messages on the bus, and thus, to detect these malicious messages, time-based intrusion detection systems (IDSs) have been proposed. However, time-based IDSs are usually trained and tested on low-fidelity datasets with unrealistic, labeled attacks. This makes difficult the task of evaluating, comparing, and validating IDSs. Here we detail and benchmark four time-based IDSs against the newly published ROAD dataset, the first open CAN IDS dataset with real (non-simulated) stealthy attacks with physically verified effects. 
We found that methods that perform hypothesis testing by explicitly estimating message timing distributions 
have lower performance than methods that seek anomalies in a distribution-related statistic. 
In particular, these ``distribution-agnostic'' based methods outperform ``distribution-based'' methods by at least $55\%$ in area under the precision-recall curve (AUC-PR). 
Our results expand the body of knowledge of CAN time-based IDSs by providing details of these methods and reporting their results when tested on datasets with real advanced attacks. Finally, we develop an after-market plug-in detector using lightweight hardware, which can be used to deploy the best performing IDS method on nearly any vehicle. 

\end{abstract}

\section{Introduction} \label{Introduction}

Modern vehicles are commonly drive-by-wire. That means that the states of vehicle subsystems are constantly being updated through broadcast messages between small dedicated computers called \textit{electronic control units} (ECUs). 
With the increasing connectivity between ECUs, inherent security risks are becoming evident. 
Controller Area Network (CAN) is the default protocol used in the automotive industry~\cite{bosch1991CAN}. 
It reduces wiring complexity by allowing ECUs to communicate on a common channel or bus with a standard protocol. 
Due to the continuously growing attack surface and the lack of authentication and lack of encryption, CAN is one of the main means by which adversaries can attack vehicles. 

Vehicle complexity increases the attack surface by allowing the execution of (remote) attacks that have been shown to be life threatening. 
Examples include the injection of malicious messages through cellular networks with the purpose of taking control of targeted vehicles~\cite{checkoway2011comprehensive, nie2017free} and the demonstrated attacks by Miller and Valasek causing unintended acceleration, deactivation of vehicles' brakes, as well as turning the steering wheel~\cite{miller2015remote, miller2016can}. 
In addition to these remote attacks, direct access to the CAN bus is generally easy to obtain, and for research purposes, most CAN attacks are administered via adding a node (ECU) through a the standard on-board diagnostic (OBD) II port.


\subsection{Preliminaries}
CAN messages or frames have two main parts: an arbitration ID (AID), typically composed of 11 bits, and a data field, composed of up to 64 bits. 
The AID functions as a label for the frame and serves to determine, by a process called arbitration, which message is received by the CAN bus when messages are transmitted simultaneously. 
The data communicates the current state of a vehicle system---for example, the anti-lock braking system. 
Commonly, messages with the same AID are sent regularly---most AIDs occur at their fixed frequency with minor aberrations---and often with redundant data. 

CAN attacks have been classified using a three-tiered taxonomy: fabrication, suspension, and masquerade attacks~\cite{cho2016fingerprinting}. 
Here we focus on fabrication attacks that, by definition, add a node (ECU) to the CAN bus and transmits messages  with the intention of overloading the bus or overwriting messages to manipulate vehicle functionality. 
While these are the simplest attacks to execute, their effect on vehicles has been repeatedly proven~\cite{song2016intrusion, moore2017modeling}. 
Notably, adding messages to the bus changes the time gap between subsequent messages, which leads to our focus---intrusion detection systems (IDSs) targeting fabrication attacks via identifying message timing anomalies.

Due to the redundancy of CAN messages, to ``overwrite'' a particular AID's message in a fabrication attack (\textit{Targeted ID Attack}), an attacker must either broadcast messages   at a rate faster than the true message \textit{(flooding delivery)} or directly after the true message with the target ID is sent \textit{(flam delivery)} \cite{ROADdataset}. 
A targeted ID attack using flam delivery is the most stealthy kind of fabrication attack, as the resulting timing perturbation is relatively minor, and thus provide the most difficult fabrication attack testing for time-based detection methods. 
Denial of service (DOS) and fuzzing attacks use flooding delivery with the goal of overloading the bus rather than targeting a particular ID, resulting in less stealthy fabrication attacks.

    \vspace{-0.5em}

\subsection{Related Work} 
\label{sec:related-work}
After early vehicle security work of Hoppe et al.~\cite{Hoppe_Kiltz_Dittmann_2009}, who note that monitoring for increased message frequency poses a potentially effective CAN IDS, initial research began to prove the concept that fabrication attacks can be accurately identified via message timing anomalies.  
Following works considered simple heuristics based on the counts of messages in a time window \cite{Otsuka_Ishigooka_Oishi_Sasazawa_2014} or the average observed time gap between messages
\cite{Gmiden_Gmiden_Trabelsi_2016, song2016intrusion, moore2017modeling}. 
Later works began applying more sophisticated statistical machinery to the problem of unsupervised fabrication attack detection via timing-based methods:
Tomlinson et al. \cite{tomlinson2018} consider two online methods and one pre-trained unsupervised method, all based on the mean inter-message time gap; 
Hamada et al. \cite{Hamada_Inoue_Ueda_Miyashita_Hata_2018} train Gaussian mixture models to identify 
time gaps in the future that occur with low probability; 
Olufowobi et al. \cite{Olufowobi_Bloom_Young_Zambreno_2018} create a complex simulation for message timing from CPU scheduling research, which unfortunately exhibits low precision (many false alerts); 
Kuwahara et al. \cite{Kuwahara_Baba_Kashima_Kishikawa_Tsurumi_Haga_Ujiie_Sasaki_Matsushima_2018} create vectors of message counts in a time window, then use a z-score on nearest neighbor distance to find anomalous windows;
Han et al. \cite{Han_Kwak_Kim_2018} test survival analysis techniques; 
Olufowobi et al. \cite{Olufowobi_Ezeobi_Muhati_Robinson_Young_Zambreno_Bloom_2019} use CUMSUM change detection; 
Young et al. \cite{Young_Olufowobi_Bloom_Zambreno_2019} apply the Fast Fourier Transform (FFT) to a square wave representation of messages and find anomalies in the frequency domain representation;
Avatefipour et al. \cite{Avatefipour_Al-Sumaiti_El-Sherbeeny_Awwad_Elmeligy_Mohamed_Malik_2019} test a combination of a modified bat algorithm with a one-class support vector machine (OCSVM). 

Noteworthy takeaways from the survey above include:
\begin{itemize}
\item[\textbf{T1}] Very simple methods, e.g. see Moore et al. \cite{moore2017modeling} report strong detection results as do more sophisticated methods, e.g., Olufowobi et al. \cite{Olufowobi_Ezeobi_Muhati_Robinson_Young_Zambreno_Bloom_2019}. \\
\item[\textbf{T2}] Methods that process a time window of data can, depending on implementation,  introduce a lag in detection (of up to the length of the time window) and an inability to distinguish specific inter-message time gaps that are anomalous \cite{Otsuka_Ishigooka_Oishi_Sasazawa_2014, Hamada_Inoue_Ueda_Miyashita_Hata_2018}. \\
\item[\textbf{T3}] Few of $\smalltilde 20$ proposed methods are evaluated on real (non-simulated), public data, as appropriate public CAN datasets were not available at the time of the earlier research, although a few later works (e.g., \cite{Olufowobi_Ezeobi_Muhati_Robinson_Young_Zambreno_Bloom_2019, Kuwahara_Baba_Kashima_Kishikawa_Tsurumi_Haga_Ujiie_Sasaki_Matsushima_2018})  use now-public datasets. \\
\item[\textbf{T4}] Most of the works fail to cite many directly comparable previous works. 
\end{itemize}

\subsection{Motivation \& Contributions}
The newly released ROAD Dataset \cite{ROADdataset} provides real (non-simulated), physically-verified, CAN data with labeled fabrication attacks, in particular many using the stealthy flam delivery. 
In light of \textbf{T1} and \textbf{T3}, 
we leverage this opportunity to benchmark some intuitive, straightforward approaches to time-based detection of fabrication attacks.  
The methods chosen aim to identify and compare the  statistical concepts foundational to time-based CAN IDS methods on a high fidelity dataset. 
To the best of our knowledge, an analysis benchmarking several different time-based IDS methods has not yet been thoroughly performed.



The contributions of this research are summarized as follows. 
In light of \textbf{T4}, we provide a thorough survey of the previous time-based CAN IDS research targeting fabrication attacks. 
Second, we provide definitions of four different time-based IDSs, 
stating the conditions needed to raise alarms in each of the methods. 
We implement the detailed methods, enabling their further deployment and comparison. 
Third, we test the implemented methods on a publicly available CAN IDS dataset, ROAD \cite{ROADdataset}. In contrast with previous work in the area, the test set includes many fabrication attacks that were administered to a real vehicle's CAN with effect to the vehicle physically verified; notably, this dataset is publicly available and does not contain simulated data.  
Fourth, we compare the proposed time-based IDS methods based on their detection ability. 
Lastly, we discuss implementation details and the operation of the best performing method in an OBD II plugin. 
In light of \textbf{T2}, this includes detection latency analysis of our Binning detector.


\section{Methods} 
\label{Methods}

In this section, we define the threat model and dataset used throughout the paper (Sec.~\ref{Methods:Threat Model}), the detection methods that we compared (Sec.~\ref{Methods:Detection Methods}), an analysis of detection latency (Sec.~\ref{Methods:Detection Latency}), and the accuracy metrics we used for comparison (Sec.~\ref{Methods:Performance Comparison}).

\subsection{Threat Model \& Dataset} 
\label{Methods:Threat Model}
 \label{Methods:Datasets}
Cho and Shin~\cite{cho2016fingerprinting} provide a widely known CAN attack terminology. Specifically, they used the term \emph{weakly compromised ECU} to denote an ECU that an adversary is able to silence by suspending message transmission. Conversely, they used the term \emph{fully compromised ECU} for an ECU over which an adversary has complete control, including the ability to send fabricated messages and gain memory access. 


In this work, we used the ROAD dataset \cite{ROADdataset} that involve a fully compromised ECU introduced to the CAN bus using the OBD-II port. 
The ROAD Dataset is the first open dataset with real (non-simulated), stealthy (using flam delivery) fabrication attacks that have physically verified effects on the vehicle. 
These characteristics  make the ROAD dataset ideal for benchmarking time-based IDS methods. 


The ROAD dataset consists of 12 ambient captures (log files) containing about three hours of ambient (non-attack) data and 33 attack captures that last in total about 30 minutes. Table \ref{Table: ids logs} lists contents of attacks and logs in the ROAD dataset, and indicates the subset of logs used in this paper. The ROAD dataset contains several types of attacks (see Table \ref{Table: ids logs}, bottom): (1) fabrication attacks, including fuzzing attacks and several different targeted ID attacks using flam delivery; (2) masquerade attacks and (3) an advanced ``Accelerator'' attack. Note that (2) and (3) do not alter timing characteristics and are thus out of scope for this paper. 
For thorough testing, several of the attacks are run and logged more than once. 
The ROAD data was obfuscated 
to ensure anonymity of the vehicle while 
preserving aspects that are needed for testing IDS methods. 
For more details on the data collection and obfuscation procedure, 
refer to the work by Verma et al.~\cite{ROADdataset}. 

Below we describe the specifics of the data we used from the ROAD dataset for the training and testing phases.

\begin{table}[t]
\vspace{-1.7em}
\centering
\caption{Logs in ROAD CAN Intrusion Detection Dataset.}
\label{Table: ids logs}
\tabcolsep=0.1cm
\renewcommand{\arraystretch}{1.1}
\begin{tabular}{| c | l c c r |}
\hline
& \textbf{Description} & \textbf{\# Logs} & \textbf{Used} & \textbf{Duration (min)} \\
\hline
\parbox[t]{2mm}{\multirow{3}{*}{\rotatebox[origin=c]{90}{Training}}} & Dynamometer Various Ambient & 10 & \cmark & 108.2 \\
& Road Various Ambient & 2 & \xmark & 70.6 \\
\cline{2-5}
& \emph{\textbf{Total}} & 12 & 10 & 178.8 \\
\hline
\parbox[t]{2mm}{\multirow{14}{*}{\rotatebox[origin=c]{90}{Testing}}} & Accelerator Attack (In Drive) & 2 & \xmark & 2.7 \\
& Accelerator Attack (In Reverse) & 2 & \xmark & 3.2 \\
& Correlated Signal Fabrication Attack & 3 & \cmark & 1.3 \\
& Correlated Signal Masquerade Attack & 3 & \xmark & 1.3 \\
& Fuzzing Fabrication Attack & 3 & \cmark & 0.7 \\
& Max Engine Coolant Temp Fabrication Attack & 1 & \cmark & 0.4 \\
& Max Engine Coolant Temp Masquerade Attack & 1 & \xmark  & 0.4 \\
& Max Speedometer Fabrication Attack & 3 & \cmark & 3.9 \\
& Max Speedometer Masquerade Attack & 3 & \xmark & 3.9 \\
& Reverse Light Off Fabrication Attack & 3 & \cmark & 2.1 \\
& Reverse Light Off Masquerade Attack & 3 & \xmark & 2.1 \\
& Reverse Light On Fabrication Attack & 3 & \cmark & 3.2 \\
& Reverse Light On Masquerade Attack & 3 & \xmark & 3.2 \\
\cline{2-5}
& \emph{\textbf{Total}} & 33 & 16 & 28.4 \\
\hline
\end{tabular}
\vspace{-1.5em}
\end{table}

\subsubsection{Training}
\label{sec:training}
Training or ambient data in the ROAD dataset was collected when the vehicle was on the dynamometer or on the road while performing diverse, potentially unusual but benign, driving activities (e.g., unbuckling seatbelts or opening doors while driving). 
In this work, for training, we used all 10 data logs that were generated on the dynamometer only (see Table \ref{Table: ids logs}, top). We select only these ambient dynamometer logs since all logs containing attacks, which make up our test data, were collected on a dynamometer. 

\begin{figure}[!htbp]
\vspace{-0.5em}
\setlength\abovecaptionskip{-0.3\baselineskip}
\centering
\includegraphics[width=1.0\columnwidth]{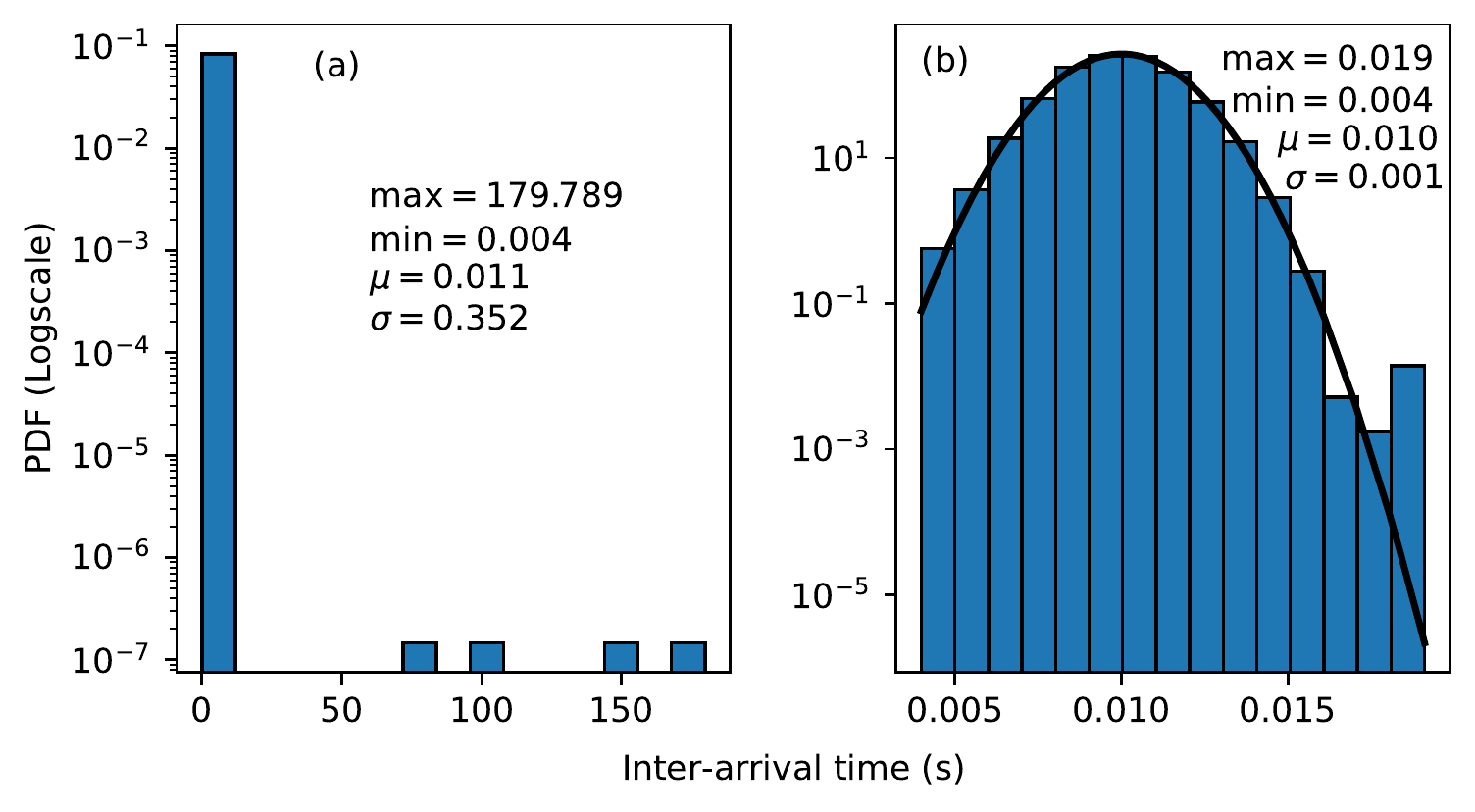}
\caption{Probability Density Function (logscale) of inter-message time distributions for AID \texttt{0xD0} in the training dataset (a) with outliers and (b) without outliers and with the Gaussian fit of the PDF shown in black. Note that without removing outliers, data is not Gaussian.}
\label{Fig: Training distribution AID 208}
\label{Fig: Training PDF AID 208}
\vspace{-1em}
\end{figure}

The time-based IDS methods that we benchmark in this work rely on analyzing inter-message time distributions of each AID. 
Thus, characterizing inter-message time distributions is an important step during the training phase. 
Fig.~\ref{Fig: Training distribution AID 208} shows the inter-message time distribution of AID \texttt{0xD0} (a targeted AID in one of the attacks). We annotate the figure with the values of the mean $\mu$, standard deviation $\sigma$, minimum value $\min$, and maximum value $\max$. 
Fig.~\ref{Fig: Training distribution AID 208}(a) shows the distribution without removing outliers in training. 
We notice that although most inter-message times are short (contained in the bar with almost six orders of magnitude), there are a few outliers corresponding to events transmitted at asynchronous rates (more than 50 seconds). 
Given that these outliers have the potential to skew the inter-message time distributions, we test results with and without outliers. 

To remove outliers, we used the Minimum Covariant Determinant (MCD) method~\cite{rousseeuw1999fast}. 
MCD is an  estimator of multivariate location (mean) and scatter (covariance) that is  robust to outliers in the data and for which a fast algorithm is available. 
MCD serves as a convenient and efficient tool for identifying and omitting outliers while fitting a Gaussian to the remaining inliers. 
For removing outliers, we used the EllipticEnvelope class in \texttt{SciKit-Learn} \cite{scikit-learn}.
We set the contamination argument to $0.01\%$ to control the proportion of expected outliers. 
Fig.~\ref{Fig: Training distribution AID 208} shows the inter-message time distribution (a) with and (b) without outliers in training. 
Clearly data in (a) (with outliers) is non-Gaussian;  in (b) (without outliers), the Gaussian is plotted in black over the data histogram. 

We observe that removing the outliers significantly shrinks the domain of the random variable ($\sigma=0.352, \max=179.789$ with outliers versus $\sigma=0.001,\max=0.019$ without). 
By removing outliers, we believe we can ensure a better characterization of the inter-message time distribution, and thereby obtain a better detector. 
We test this hypothesis by comparing results with and without outliers in Sec.
\ref{Results}.


\subsubsection{Testing}

We used all 16 available log files for testing time-based detectors (namely, those containing fabrication attacks): the fuzzing attacks and the five different fabrication targeted ID attacks (see Table \ref{Table: ids logs}, bottom). 
We aggregated all of these logs for evaluating the performance of the time-based IDS methods. 
Note that we did not remove outliers in the testing dataset.
Importantly, among the 16 logs files that we used for testing, there was a total of 1,588,263 messages, of which 61,516 were attacks. This means that the proportion of positive samples is about $3.9\%$, which makes the testing dataset highly imbalanced. We revisit the imbalanced effect of the testing dataset in Sec.~\ref{Results}.

\subsection{Detection Methods} 
\label{Methods:Detection Methods}
We tested four detection methods that exploit the timing regularities of CAN messages: Mean Inter-Message Time (Sec.~\ref{Methods:MIAT}), Binning (Sec.~\ref{Methods:Binning}), Fitting a Gaussian Distribution (Sec.~\ref{Methods:Gaussian}), and Kernel Density Estimation (KDE) (Sec.~\ref{Methods:KDE}). 
While the first two rely on heuristics based on the inter-message times, the latter two follow previous anomaly detection works by fitting a continuous distribution to the inter-message times and detecting time gaps with low $p$-value \cite{ bridges2017setting}. 
We employ the two-sided $p$-value (of a sample $x$ given a probability distribution $f$), denoted by $pv_{f}$, where $P_f$ is the probability with respect to $f$ as:
\begin{equation}
    \label{eq:pv} 
    pv_{f}(x) := \int_{\{t:f(t)\leq f(x)\}} f\,\, dt = P_{f}\left(\{t:f(t)\leq f(x)\}\right).
\end{equation}

All methods except for Binning use a two-step process for determining if a message is malicious. 
\textbf{Step (1) Comparison:} this step identifies whether a message is suspicious by comparing it to a threshold identified 
during training. 
\textbf{Step (2) Check for sufficiency:} this step determines whether the frames identified as suspicious in Step (1) are sufficiently anomalous to warrant an alert. This is done by checking if suspicious inter-message times occur repeatedly; the intuition is that to manifest physical changes, fabrication attacks must have repeated injections to continually overwrite the ambient messages. 
If sufficiency is confirmed, the messages are marked as malicious.


In this section, let $X_{\text{AID}}$ be the set of inter-message times for a particular AID in the ROAD dataset and let $x$ be an arbitrary inter-message time in $X_{\text{AID}}$. Again, $\mu$ denotes the average inter-message time for this AID. We run each method using a range of detection thresholds $\alpha$, where $\alpha$ is defined below for each.

\subsubsection{Mean Inter-Message Time (Mean)}
\label{Methods:MIAT}
The first method uses the empirical mean $\mu$ calculated during training and a given threshold $\alpha\in [0,1]$. 
\textbf{(1) Comparison}: consider if inter-message time $x\leq \alpha \mu$. \textbf{(2) Check for  sufficiency}: if $x\leq \alpha \mu$ holds for three out of the last six inter-message times, the frame is labeled malicious. 
We compute results for thresholds  $\alpha_i = i/18$ for $i = 1, \dots, 18$. 
We use 18 thresholds to be  consistent with the distribution-based methods (\ref{Methods:Gaussian}, \ref{Methods:KDE}).  


\subsubsection{Binning}
\label{Methods:Binning} 

The second method considers the number
of messages transmitted during a certain window of time.
We let $\alpha_i=(2+i)/2$ for $i=1,\ldots,18$.
In order to identify the appropriate threshold, for each $\alpha_{i}$, we labeled a
message malicious when the last six frames arrived in less than $\alpha_{i} \mu$ seconds. As intuition, note that in a time window of length $\alpha\mu$ we expect $\alpha$ messages to occur. Hence for $\alpha<6$, this method will alert when extra messages are occurring.




\subsubsection{Fitting a Gaussian \text{Distribution} (Gaussian)}
\label{Methods:Gaussian}
The third method  fits a Gaussian distribution, $f_G$, to the inter-message times $x$ (regardless of the shape of the data) and computes the two-sided $p$-value defined in Eq. \ref{eq:pv}. 
Let $ S:=\{0.001,\dots, 0.009\}\cup  \{0.01, \dots ,0.09\}.$ 
Detection was run for each $
\alpha \in S$ as follows. \textbf{(1) Comparison}: for each $x\in X_{\text{AID}}$, compute $pv_{f_G}(x)$ (as in Eq. \ref{eq:pv}). \textbf{(2) Check for sufficiency}: if $pv_{f_G}(x)\leq \alpha$ for three consecutive inter-message times, the frame is labeled malicious. 




\subsubsection{Kernel Density Estimation (KDE)} \label{Methods:KDE}
The final method uses a KDE using a Gaussian kernel to establish a probability density function, $f_{\text{KDE}}$, of the inter-message time data. 
Steps for detection follow the Gaussian method's \textbf{(1)} and \textbf{(2)} with $pv_{f_{\text{KDE}}}$ instead of $pv_{f_G}$. 
Detection results are computed for each $\alpha \in S$ as in the Gaussian method (\ref{Methods:Gaussian}). 



\subsection{Detection Latency} \label{Methods:Detection Latency}
For all methods except Binning, each frame is scored and alerted; hence, latency is simply the computational time for the detection. 
Note that the Binning method, by design does not compute counts for moving time windows, but instead stores the previous five message times and asks if the next message occurs too soon, specifically if the last six messages occurred in less than $\alpha \mu$s. 
Hence, detection latency depends on $\alpha.$ 
Here we suppose that there is at least one illegitimate frame sent between each pair of ambient  frames. 
If $\alpha \geq 5 $ detection is immediate (simply computation time for the algorithm) as the previous five message (all legitimate) will require $\smalltilde 4\mu$ s; thus, the time gap from the first initial illegitimate frame to the sixth previous will be under $5\mu$. 
Similarly, the time gap from the second illegitimate frame to the sixth previous is at most $4\mu$, meaning for $\alpha \in [4,5)$ detection latency is $2 \mu$ plus computational time. 
Finally, the time gap from the third or any subsequent illegitimate frame to the sixth previous message is at most $3\mu$. 
This means our detection latency will be at most $3\mu$ plus computation time for $\alpha <4.$
Note that detection accuracy may vary greatly among different values of $\alpha$. 
To put this in context, most AID frequencies are on the order of 10Hz to 100Hz, so the latency is a fraction of a second for all thresholds $\alpha$.  



\subsection{Detection Ability Comparison} \label{Methods:Performance Comparison}
The basis for the comparison between the time-based IDS methods is based on counting the number of CAN messages that were labeled as true positives (TP), false positives (FP), false negatives (FN), and true negatives (TN). Several important classification metrics are based on these numbers: \textit{Precision}, defined as $\frac{TP}{TP+FP}$, gives the likelihood that a detected message is an attack; \textit{Recall}, defined as $\frac{TP}{TP + FN}$, gives the likelihood that an attack is detected. Since higher precision often comes at the price of higher recall (and vice versa), it is important to consider a balance of both metrics, and the standard balanced metric is the \textit{F1 score}, defined as $2 \times \frac{precision \times recall}{precision + recall}$. 

To compare the performance of these time-based methods over a range of detection thresholds ($\alpha$), we consider the \textit{Precision-Recall (PR) curve}, which plots precision against the recall at different thresholds. We choose to use the PR curve rather than the more commonly used ROC curve, which is less suited to imbalanced datasets~\cite{davis2006relationship} 
(recall  that  the  proportion  of  attack  messages,  i.e.,  positive samples,  is  about 3.9\%). This PR curve illustrates how the accuracy of a method changes at different performance thresholds, and the optimal threshold in terms of F1 score will generally appear towards the furthest top-right point of the curve. In our evaluation, we determine the optimal threshold to be the threshold that results in the highest F1 score. 

This curve is also provides an important metric for evaluating a methods head-to-head independent of a specific threshold, namely, the area under the PR curve (AUC-PR), where a higher percentage indicates better overall performance. Note that is is important to consider this threshold-independent metric, since a true unsupervised method cannot consult the test data to determine an optimal threshold. Note that the \textit{baseline} of the PR curve is the proportion of positive samples $\frac{TP + FN}{TP + FN + TN + FP}$.


\section{Results} 
\label{Results}

\begin{figure}[t]
\setlength\abovecaptionskip{-0.5\baselineskip}
\centering
\includegraphics[width=1.0\columnwidth]{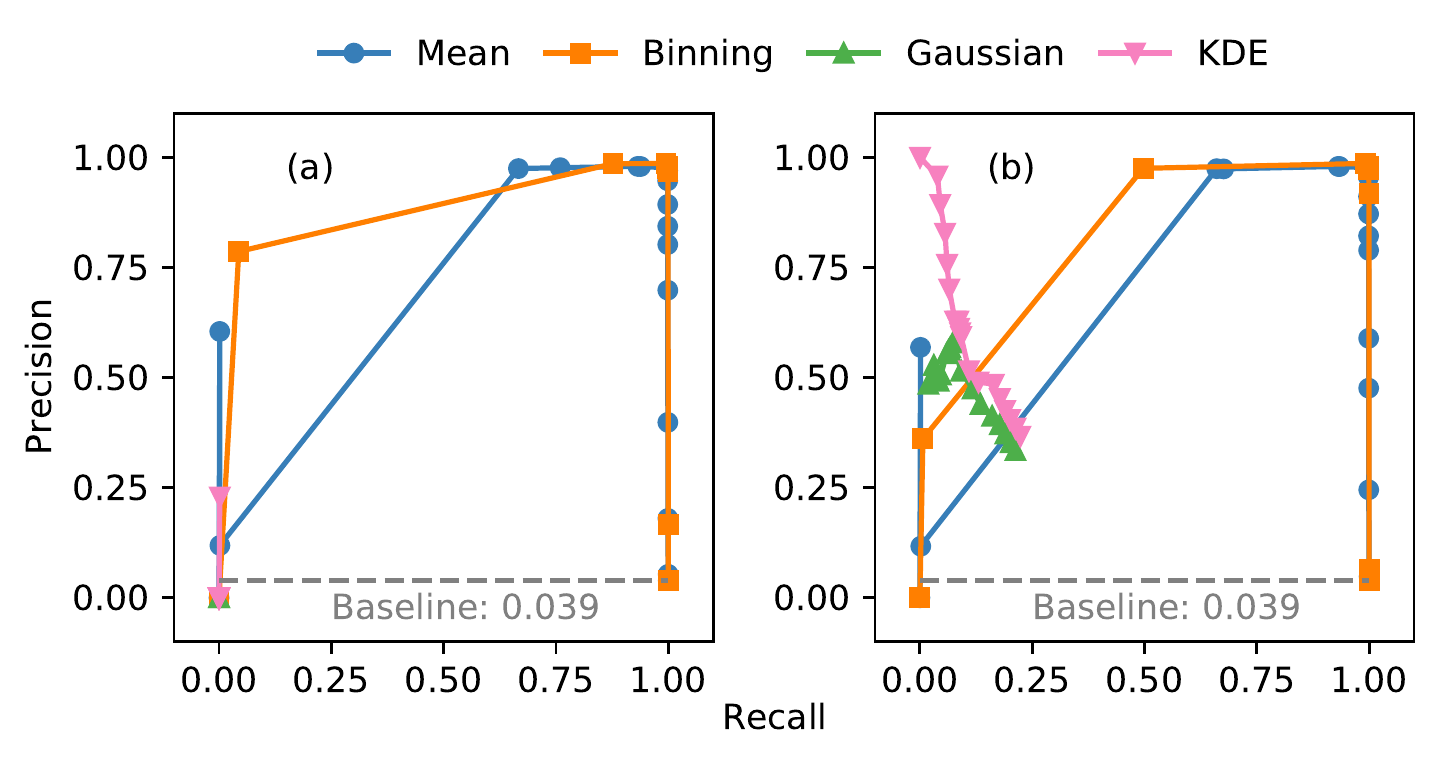}
\caption{Performance comparison based on PRC. (a) with (b) without outliers. The Binning method performs best overall.}
\label{Fig: complete PRC}
\end{figure}

For each method, the PR curve is shown in Fig. \ref{Fig: complete PRC}
and the AUC-PR is given in Table \ref{Table: performance comparison}.  

We first consider the method performance independent of a detection threshold. We report the performance comparison between the four methods across every threshold using AUC-PR in Table~\ref{Table: performance comparison}, and highlight the method with best performance in bold. 
Overall, we find that methods that rely heavily on the distribution (i.e., Gaussian and KDE methods) tend to perform much worse and be significantly affected by outliers, as opposed to methods that do not have these distributional assumptions (i.e., Mean and Binning methods). With outliers present, the AUC-PR for both the Gaussian and KDE method are near 0, and while they perform significantly better when these outliers are removed from the training set, both still have a very low AUC-PR of under 15\%. 
Removing outliers also makes the performance of the Mean method slightly better (from 68.90$\%$ to 69.05$\%$) but interestingly, it worsens the performance of the Binning method (from 87.63$\%$ to 82.18$\%$). 
The Binning method has the best performance both with and without outliers in training, achieving over 82\% in both cases. 

We next consider the performance of the methods at different threshold values. This is done illustrate how these thresholds affect the methods and determine the optimal threshold and resulting F1 score for each. 
Fig.~\ref{Fig: complete PRC} gives a more detailed illustration of the conclusions given above: regardless of the presence of outliers, both Gaussian and KDE methods tend to perform poorly at all detection thresholds, while the Mean and Binning methods perform much better, particularly at certain thresholds (indicated by the blue and orange points near $(1.00,1.00)$). 

For the distribution-agnostic methods, we found that the optimal threshold and resulting optimal F1 score were essentially the same with or without outliers present in training. We found that the optimal threshold value for the Mean method is $\alpha = 0.44$, At this threshold, the value of the F1 score is $\smalltilde98.6\%$ (precision is 97.6\% and recall is 99.6\%). 
This means that reporting (as anomalies) messages that are less than $44\%$ of the expected inter-arrival time mean (per AID) produces the best tradeoff between precision and recall. 
We also found that for the binning method, the optimal value for the threshold $\alpha$ (for this method, the window length) is 3.5. At this threshold, the F1 score is $\smalltilde99.0\%$ (precision is 98.6\% and recall is 99.2\%). 
This means that reporting messages (as anomalies) that do not resemble the expected number of messages in a window of length of 3.5 produces the best tradeoff between precision and recall. 

For the distribution-based methods, we consider only the case in which outliers are removed, since these methods are essentially useless without this preprocessing step. 
The optimal threshold (the p-value) for the Gaussian and KDE methods was $0.09$. At this value the F1 score achieved is $25.99\%$ and $27.79\%$, respectively.



\begin{table}[t]
\centering 
\caption{AUC-PR comparison with and without outliers in training. The Binning method (in bold), is the best performing method in both cases.}
\label{Table: performance comparison}
\begin{tabular}{| l | r | r |}
\hline
\textbf{Method}  & \textbf{With Outliers} & \textbf{Without Outliers} \\
\hline
Mean             & 68.90\%                            & 69.05\%                               \\
\textbf{Binning} & \textbf{87.63\%}                   & \textbf{82.18\%}                      \\
Gaussian         & 0.00\%                             & 8.99\%                                \\
KDE              & 0.02\%                             & 13.88\%  \\ 
\hline
\end{tabular}
\end{table}

\subsection{OBD-II Plugin Prototype} \label{Results: OBD-II Plugin Prototype}
To prove the proposed detector is a viable vehicle-agnostic detector able to run on small edge computing devices efficiently, we implemented the Binning method (Sec \ref{Methods:Binning}) on an  OBD-II  plugin. 
We devised an edge computing device using a Raspberry Pi 3B+ with Raspbian Buster  providing 1GB of RAM, a 1.4GHz ARMv8 processor, and Python support coupled with an interface board, Industrial Berry's  CANBerry Dual 2.1 \cite{can_berry} to facilitate CAN communication.
Once attached to the CAN bus (plugged into an OBD II of a running vehicle), the device runs an algorithm to identify the CAN's pre-configured bitrate along with other metadata, e.g., the vehicle identification number (VIN).  
The device then automatically collects CAN data from the vehicle and trains the parameters for each AID's detector. 
At this point the driver would ideally exercise as much of the vehicle's functionality as possible in order to determine the timing characteristics of the most possible AIDs. 
Once trained, the device can deploy the time-based detectors in real time and receive alerts visually via LEDs, with metadata and logs of the alerts stored on the prototype. 
The stages of the process are started/stopped with a switch on the device. 
The  implementation into hardware follow our previous work \cite{verma2020cand}. 
We provide a demonstration video of the hardware running the Binning detector on a passenger vehicle while implementing an attack online (\url{https://www.youtube.com/watch?v=mQxkycW0mV4}). 

\vspace{-1.0em}
\section{Discussion} \label{Discussion}
Recalling the takeaways from our survey of the related literature, Sec. \ref{sec:related-work}, specifically,  \textbf{T1} (simple detectors have garnered strong results) and \textbf{T4} (few previous works use real, public, physically-verified, high-fidelity attack datasets for testing), we formally define, implement, comparatively evaluate four different time-based intrusion detection system (IDS) methods on the ROAD Dataset~\cite{ROADdataset}, a publicly available CAN dataset with labeled attacks. 
The Mean method (\ref{Methods:MIAT}) represents the many initial works that use heuristics based on differences from the average inter-message timing, while the Binning method (\ref{Methods:Binning}) represents the early methods that count messages in a time-window. 
As these both rely heavily on the learned mean inter-message time, a logical next step is to leverage also the second moment, variance, 
specifically, fitting a Gaussian to training data and detecting events in the tail (\ref{Methods:Gaussian}). 
Peeking at the training data (Fig. \ref{Fig: Training distribution AID 208}) we note that outliers may ruin the Gaussian fit. 
Alternatively, one may argue that non-Gaussian data should be fit with a more flexible class of  distributions. 
To this end we also tested an analogous detector based on the very flexible kernel density estimate (KDE) method (both with and without outliers in training) to test this hypothesis.

Unsurprisingly considering previous works, the two simple, distribution-agnostic  (Mean, Binning) achieve strong results (F1 scores of 0.986, 0.990, respectively). 
Moreover, our results with and without outliers in the training data show the methods are fairly robust to outliers as their results are  not substantially affected. 

Perhaps shockingly, the two distribution-based methods pale in comparison. Our analysis of results when fitting these distributions to training data with and without outliers shows that these poor detection results exhibit only minor improvements (F1 scores of 0.260, 0.278, respectively); hence, they are not amended by this pre-processing maneuver. 
Rather, it seems that the tails of the Gaussian (and KDE, whose tails are near identical) are insufficient to characterize these. In the case of the Gaussian, these result suggest that considering the inter-message time as a fraction of the mean is an informative statistic but as a $z$-score is not, perhaps because the variance is so small in these distributions. 
As the tails of the Gaussian and KDE distributions are similar, this provides intuition for the poor results of these distributional methods.

Overall, these results suggest that counting methods are perhaps slightly better than those that consider inter-message timing, and more dramatically, the heuristic/distribution-agnostic methods far outperform those that explicitly seek tails of an estimated  distribution. Inspecting the testing data reveals an artifact of flam attacks (that send one illegitimate message just after a legitimate  message); i.e., they force the attack inter-message times to be a bi-modal distribution.

The following are some limitations of our study. 
First, we do not consider 
CAN messages without fixed timing. 
Second, the ROAD dataset only includes data from a single car. It is clear from the research that these techniques will succeed on a wide variety of vehicles, but specifics, e.g., the optimal thresholds found via our test data, may or may not translate to other vehicles. 
Third, as these are unsupervised methods, how to set the threshold with no attack data is a real-world need that we do not address. 
Finally, the methods we tested are exclusively of a time-based nature and the scope of our paper is limited to fabrication attacks. 
These methods are not intended to detect attacks that do not alter message frequencies, e.g., masquerade attacks. 
For detecting such anomalous event messages, methods should also take into account changes in the payload (e.g., see~\cite{hanselmann2020canet}).


To the best of our knowledge, the results from this research show for the first time a systematic comparison of time-based detection methods on an open CAN IDS dataset with verified real-world attacks. 
While this may seem like a solved problem we note that the results of the Binning and Mean detectors (and likely most all previous works) are in fact insufficient when put in context. 
Referring to the Dataset Section (\ref{Methods:Datasets}), the test data was $\smalltilde 12.3$ minutes of driving data comprised of $\smalltilde 1.6$M CAN messages. 
With the optimal Binning detector's precision of 98.6\%, this still produces approximately $0.014\times 1.6/12.3$M = 1,821 alerts per minute.
In short, systematically pushing the state of the art in this area is still needed, and we hope this benchmark of the straightforward approaches in the area helps. 
In summary, our results indicate that future researchers may wish to focus on counting algorithms that can increase precision at (ideally) no expense to recall, and statistical machinery that relies on distributional estimates.

\small
\bibliographystyle{IEEEtranS}
\bibliography{bibliography}

\end{document}